# Constraining the Time Interval for the Origin of Life on Earth

Ben K. D. Pearce*, Andrew S. Tupper, Ralph E. Pudritz and Paul G. Higgs

Origins Institute and Dept of Physics and Astronomy,

McMaster University, 1280 Main St, Hamilton, ON, L8S 4M1, Canada

*To whom correspondence should be addressed. E-mail: pearcbe@mcmaster.ca

## Abstract

Estimates of the time at which life arose on Earth make use of two types of evidence. First, astrophysical and geophysical studies provide a timescale for the formation of Earth and the Moon, for large impact events on early Earth, and for the cooling of the early magma ocean. From this evidence, we can deduce a *habitability boundary*, which is the earliest point at which Earth became habitable. Second, biosignatures in geological samples, including microfossils, stromatolites, and chemical isotope ratios, provide evidence for when life was actually present. From these observations we can deduce a *biosignature boundary*, which is the earliest point at which there is clear evidence that life existed. Studies with molecular phylogenetics and records of the changing level of oxygen in the atmosphere give additional information that helps to determine the biosignature boundary. Here, we review the data from a wide range of disciplines to summarize current information on the timings of these two boundaries. The habitability boundary could be as early as 4.5 Ga, the earliest possible estimate of the time at which Earth had a stable crust and hydrosphere, or as late as 3.9 Ga, the end of the period of heavy meteorite bombardment. The lack of consensus on whether there was a late heavy meteorite bombardment that was significant enough to prevent life is the largest uncertainty in estimating the time of the habitability boundary. The biosignature boundary is more closely constrained. Evidence from carbon isotope ratios and stromatolite fossils both point to a time close to 3.7 Ga. Life must have emerged in the interval between these two boundaries. The time taken for life to appear could, therefore, be within 200 Myr or as long as 800 Myr.

## 1. Introduction

When did life on Earth emerge? Not only is this a key question in understanding the origin of life on our planet, but the estimated time interval may eventually lend us insight into the frequency of life's emergence elsewhere in the Universe. Soon after the Earth formed, the collision with a Mars-sized body probably gave rise to the Moon. This cataclysm melted and reshaped the Earth's mantle. Further smaller impacts may have occurred over the ensuing 0.6 Gyr or so, each of which would have been accompanied by partial melts of the Earth's crust, perhaps culminating in a last crescendo: the Late Heavy Bombardment (LHB). The atmosphere was not the "reducing" atmosphere that figures into the early Miller-Urey experiments, but one that is composed mainly of $CO_2$ (outgassed from volcanos) and $N_2$. The UV irradiation of the planet would be highly significant as there was no oxygen or ozone layer at the time to screen the surface. This radiation can be both highly destructive for the survival of complex organics, but could also have stimulated them. It is in this violent, highly changing environment that the steps that led to life took place.

The window during which the origin of life must have occurred is demarcated by inner and outer time boundaries. We call the outer boundary the *habitability boundary*. This is our estimate of the time at which the Earth first became habitable for life. We call the inner boundary the *biosignature boundary*. This is the time of the earliest convincing evidence of life in the form of fossil and/or chemical biosignatures. In this review we aim to bring together the many different lines of evidence that allow us to estimate the times of the two boundaries. The biosignature boundary has been reviewed by Schopf (2006) and Buick (2007). The habitability boundary has been reviewed by Zahnle *et al.* (2007). More recently, information pertaining to both boundaries were reviewed in a textbook by Gargaud *et al.* (2012). They concluded that life emerged on Earth at an uncertain date between 4.3 and 2.7 Ga. Here, we bring the latest astrophysical constraints and biological signatures together to more precisely constrain this time interval for the origin of life. This review considers more recent data that reframes the entire argument for when life emerged since the previous reviews on this topic. Much of the relevant information on the timing of the origin comes from astrophysics, planetary science, geology and palaeontology. Here we aim to present these results in a way that is accessible to the broad scientific community.

In his review of the RNA World hypothesis for the origin of life, Joyce (2002) illustrated the timeline of events pertaining to the origin of life on Earth, and this has been widely used by various scientific communities. In an attempt to produce a similarly useful summary, we have illustrated the evidence reviewed in this paper in a similar way in Figure 1, emphasizing the astrophysical evidence that leads to positioning the habitability boundary and the biological evidence that leads to positioning the biosignature boundary. Table 1 gives further information and citations of the events included in Figure 1.

We begin with the astrophysical constraints on the habitability boundary, and gather the data on the age of the Solar System, Earth, and Moon, and the radiative cooling time of the Earth's liquid magma-covered surface. The habitability boundary is mainly constrained by the time at which liquid water could exist on the planet's surface. Liquid water would have been a requirement for the emergence of the kind of life that exists on Earth today. Water is often considered as a requirement for habitability because of its unique role as a universal solvent, which is why NASA has invoked a "follow the water" strategy towards searching for extraterrestrial life. We will therefore discuss the theories of cooling of the Earth's surface and atmosphere after major impact events, as these make predictions about the time at which a liquid ocean could have formed. One catastrophic impact is the collision that created the Moon, but other large impacts may have continued for a much longer period, and may have also "reset the clock" as far as habitability is concerned. The rate at which major impacts died away during the early history of the solar system and the debate about whether there was a LHB phase with a high incidence of impacts are therefore very relevant to the timing of the outer boundary. Independent evidence for the outer boundary comes from the geological record. The existence of certain minerals that form in the presence of water also gives constraints on the time by which a liquid ocean must have formed.

When we consider the biosignature boundary, the most direct evidence for the existence of life on Earth comes from the presence of fossils. These may be in the form of microfossils (*i.e.* remnants of life at the level of individual cells) or macrofossils, in particular stromatolites, which are layered structures thought to be created by photosynthetic micro-organisms. The interpretation of microfossils has been controversial, and we will discuss whether some of the earliest samples claimed to be microfossils are in fact of non-biological origin. However, the timing of the biosignature boundary does not depend too much on the uncertainty around microfossils, because indirect evidence for the presence of life in the form of carbon isotopic signatures precedes most potential microfossil ages. Furthermore, stromatolite samples are known with fairly similar dates to the earliest microfossils, and the most recently proposed stromatolite samples also go as far back as the isotopic signatures.

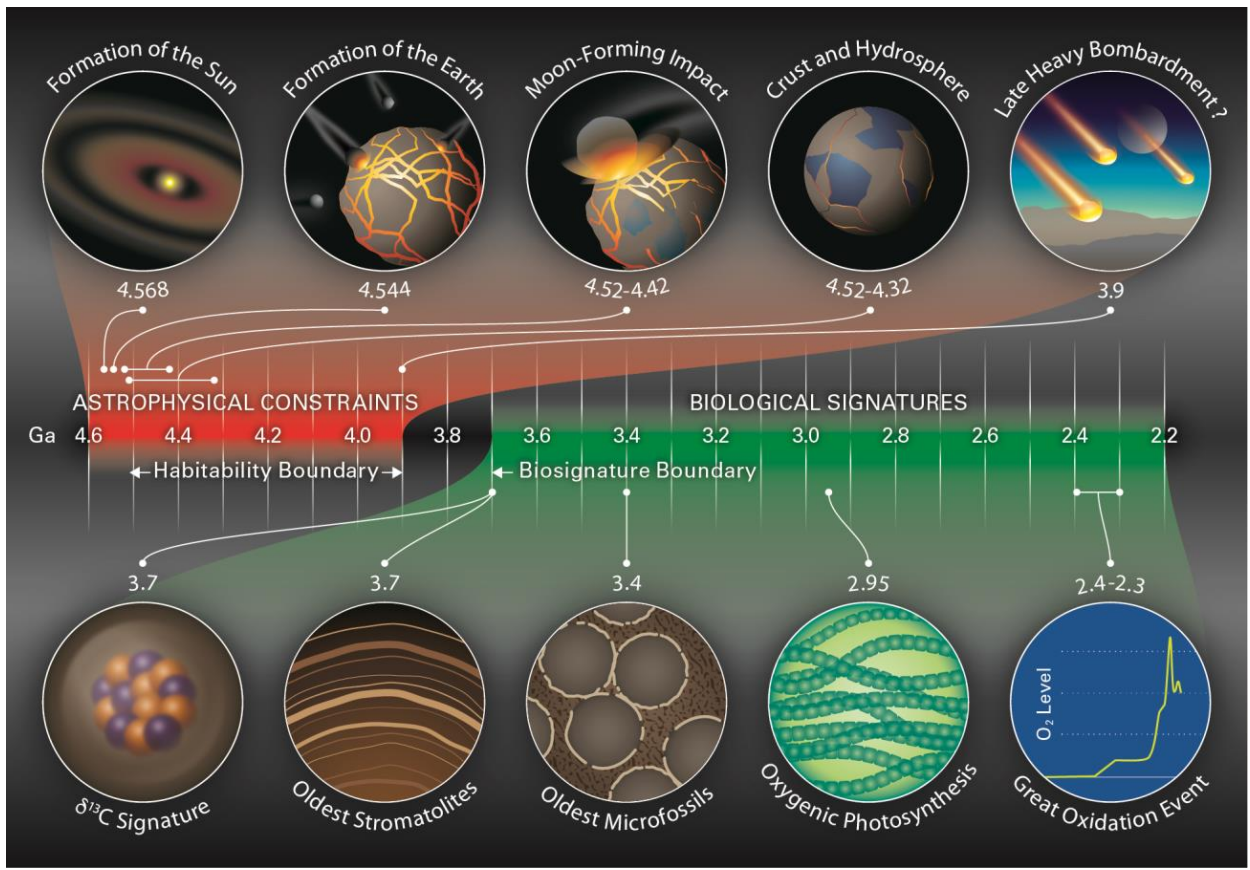

**Figure 1.** Timeline which illustrates the astrophysical constraints constraining the time of the habitability boundary and the biological signatures constraining the time of the biosignature boundary. Uncertainties about whether there was a late heavy bombardment mean that the position of the habitability boundary is still poorly constrained, while the evidence for the biosignature boundary is beginning to converge. LHB, Late Heavy Bombardment

**Table 1.** Summary and scientific basis of the events and biosignatures in the timeline.

| | Event/Biosignature | Timeline Location (Ga) | Scientific Basis | Reference(s) |
|---|---|---|---|---|
| Astrophysical Constraints | Formation of the Sun | 4.568 | CAIs in chondritic meteorites | Bouvier & Wadhwa (2010) |
| | Formation of the Earth | 4.544 | Hf-W dating | Yin *et al.* (2002) |
| | Moon-forming impact | 4.52–4.42 | Hf-W dating, Rb-Sr dating, U-Pb dating, dynamic simulations | Touboul *et al.* (2007); Halliday *et al.* (2008); Barboni *et al.* (2017); Jacobson *et al.* (2014) |
| | Formation of crust and hydrosphere | 4.52–4.32 | Numerical simulations, zircon record | Monteux *et al.* (2016); Lebrun *et al.* (2013); Zahnle *et al.* (2007); Harrison *et al.* (2005); Cavosie *et al.* (2005) |
| | Late Heavy Bombardment ? | 3.9 | Lunar crater record, lunar meteorites, dynamic simulations, physical modeling | Ryder *et al.* (2000); Cohen *et al.* (2000); Gomes *et al.* (2005); Boehnke and Harrison (2016) |
| Biological Signatures | $\delta^{13}$C signature | 3.7 | Light carbon signature in rocks of sedimentary origin | Rosing *et al.* (1999); Ohtomo *et al.* (2013) |
| | Oldest Stromatolites | 3.7 | Layered structures interpreted as stromatolites | Nutman *et al.* (2016) |
| | Oldest microfossils | 3.4 | Cell lumina and carbonaceous cell walls | Wacey *et al.* (2011) |
| | Oxygenic photosynthesis | 2.95 | Molybdenum signature | Planavsky *et al.* (2014) |
| | Great Oxidation Event | 2.4–2.3 | MIF disappears from geologic record | Sessions *et al.* (2009) |

CAIs = calcium–aluminum-rich inclusions; MIF = mass-independent fractionation.

Another form of evidence for the existence of life comes from the increase in the level of atmospheric oxygen due to the presence of photosynthetic organisms. Atmospheric oxygen is measured through geologic time using sulfur and molybdenum isotopes and transition metals in sedimentary rocks. Although a major increase in oxygen at the time of the great oxidation event (around 2.4–2.3 Ga) is well documented, it is considerably later than the likely time for the origin of life. Evidence for traces of oxygen at 2.95 Ga suggests oxygenic photosynthesis was operating quite early. This provides constraints on the time of the inner boundary that are compatible with evidence from stromatolites.

From this consolidated research, given the uncertainty in the way the LHB unfolded, we present two possible windows for the emergence of life, confined by the two boundaries discussed in this paper. To our knowledge this is the first time such a consolidation of literature has been attempted with this goal in mind.

## 2. Formation of the Solar System

### 2.1 Formation of the Sun and the Protoplanetary Disk

Observations done with a variety of telescopes (e.g. the Hubble Space telescope, the Atacama Large Millimeter Array, the Submillimeter Array) show that dusty, gaseous "protoplanetary" disks exist around all young stars (Andrews *et al.* 2016, Dutrey *et al.* 2014, Andrews *et al.* 2010). The gas phase in such disks lasts 1–10 million years so Jupiter-mass planets must have formed within that time (e.g. review by Dutrey *et al.* 2014). It was during this time that the central stars assembled by accreting material from these surrounding disks. Rocky planets were assembled somewhat later from collisions between the left-over planetesimals and sub-Mars-sized objects in the now gas-free disk (Righter & O'Brien 2011). Our solar system likely formed in this way. It all begins within dense molecular clouds in space, where even denser regions called "cores" about 0.3 light-years across form (Krumholz 2008). These cores cool and eventually collapse under their own gravity to form stars. And because they started with some non-zero angular momentum, the system collapses into a disk out of which forming stars accrete their gas (McKee & Ostriker 2007, Krumholz 2008).

The age of the Solar System can be defined as the time in which the first solid grains formed in the nebular disk around our protosun (Bouvier & Wadhwa 2010). The oldest material in the Solar System is found within chondritic meteorites, which is the most common type of meteorite found on Earth. These meteorites are rocky and undifferentiated, which is to say, their meteorite parent bodies (i.e. 100 km-sized, rocky bodies known as planetesimals) never reached high enough temperatures to melt internally. Most of the meteorites that end up on the surface of the Earth are just fragments of planetesimals that were left over from the process of planet formation. The estimate of our Solar System's age is 4.568 billion years old (Bouvier & Wadhwa 2010), and comes from the $^{207}$Pb-$^{206}$Pb isotopic dating of the oldest material within these meteorites, the so-called calcium-aluminum-rich inclusions (CAIs). $^{207}$Pb and $^{206}$Pb are the final decay products of decay chains that begin with $^{235}$U and $^{238}$U, whose half-lives are ~0.7 Gyr and ~4.47 Gyr, respectively. CAIs are submillimetre-to-centimetre-sized grains that are thought to be among the earliest solids to have formed in the protoplanetary disk. This is due to the fact that they form mostly from oxides and silicates of calcium, aluminum, magnesium, and titanium, which condense at high temperatures, and therefore would be among the first solids to form everywhere in the warm early disk (MacPherson *et al.* 2005).

### 2.2 Formation of the Earth

The earliest that life could have appeared was soon after Earth's formation. Therefore, we can slide the outer boundary for the emergence of life inwards from 4.568 Ga, to the age that the Earth formed out of planetesimals in the protoplanetary disk.

Planetary differentiation is a fundamental concept when it comes to dating the formation of the Earth. The internal composition of the Earth varies at different depths due the tendency of higher density fluids to sink under gravity (and similarly, the tendency of lower density fluids to float). During planetary accretion, the kinetic energies deposited by impacts, and the gravitational energies released from the settling of metals to the core, cause significant melting (Wood et al. 2006). The compositions of Earth's layers are therefore influenced by the cooling rate. This is what makes hafnium-tungsten ($^{182}$Hf-$^{182}$W) isotopic dating a useful tool for estimating the formation age of rocky planets in our Solar System. Tungsten is a moderate siderophile (i.e. tends to bond with metallic iron and sink to the core), whereas hafnium is a lithophile (i.e. tends to bond with oxygen, forming compounds that don't sink to the core) (Zahnle et al. 2007). $^{182}$Hf also decays to $^{182}$W with a half-life of ~9 Myr (Zahnle et al. 2007). Now, if planetary accretion and differentiation ceased before all the planet's $^{182}$Hf decayed into $^{182}$W, one would find positive $^{182}$W abundances in the primitive silicate mantle. This means one can estimate how long the planet was accreting, melting, and differentiating materials by measuring the abundance of $^{182}$W in samples of the primitive mantle. Based on models comparing the tungsten isotopic data measured in meteorites against those of the bulk silicate Earth (i.e. the primitive mantle), the time for the Earth to reach 63% completion is 11 ± 1 Myr, and to reach 90% completion is 24 Myr (Yin et al. 2002). These estimates push the outer boundary of the emergence of life on Earth to 4.544 Ga (when the Earth had formed 90% of its mass), but several other factors shift the boundary further than this.

*2.3 Formation of the Moon*

It is hypothesized that the Moon formed from a giant collision between a planetary embryo (named Theia) and the proto-Earth (Barboni et al. 2017, Young et al. 2016, Jacobson et al. 2014, Ćuk & Stewart 2012, Canup 2012, Reufer et al. 2012, Touboul et al. 2009, Halliday 2008, Touboul et al. 2007). This is made evident by the nearly identical isotopic compositions of the Moon and Earth, suggesting a vigorous mixing of materials occurred between the two bodies (Young et al. 2016). Computer simulations show how this might have occurred (Ćuk & Stewart 2012, Canup 2012, Reufer et al. 2012). The nearly identical isotopic signatures also suggest that the Moon was the last giant impact experienced by the Earth. The formation of the Moon therefore sets the initial conditions for the early Earth environment.

Similar to the bulk silicate Earth samples, tungsten isotopic abundances were measured for lunar mantle material that was collected as part of the Apollo program, and compared to meteoritic values. Models based on these comparisons estimate a date of Moon formation to be $62^{+90}_{-10}$ Myr after the formation of the Solar System (Touboul et al. 2007, 2009). Halliday (2008) used Sr isotopes of lunar rocks to constrain the Rb-Sr age of the Moon to 87 ± 13 Myr after the Solar System formed. Rb-Sr dating is based on the isotopic decay of $^{87}$Rb to $^{87}$Sr, which has a half-life of 48.8 Gyr. Most recently, Barboni et al. (2017) coupled U-Pb and Hf isotopes from lunar zircons to develop a two-stage model age for the Moon, constraining its formation to 58 ± 10 Myr after the formation of the Solar System. Lunar zircons are robust against isotopic disturbances from post-Moon formation impacts, making this model one of the most precise to date (Barboni et al. 2017).

Another line of evidence from astrophysical simulations confirms the results from isotopic dating. Jacobson et al. (2014) used dynamic simulations involving many small particles interacting under their own gravity to estimate the Moon formation age. By running several simulations beginning from a disk of planetesimals and planetary embryos, Jacobson et al. (2014) found a clear statistical correlation between the time of the last giant impact (i.e. embryo-embryo collision) and the subsequent accreted mass for Earth-like planets. They then compared the best estimate of Earth's late-accreted mass inferred from highly siderophile element abundances in Earth's mantle (e.g. Rhenium, Osmium, Iridium, Ruthenium) relative to those in chondritic meteorites to estimate the time of the last giant impact on Earth. Based on this analysis, Jacobson et al. (2014) estimate the time of the Moon-forming impact to be 95 ± 32 Myr after the formation of the Solar System. This result agrees with the results from isotopic data by Touboul et al. (2007, 2009), Halliday (2008), and Barboni et al. (2017). By making no preference between the three isotopic analyses or the dynamic simulations as methods to estimate the Moon formation age, we can slide the outer boundary for the emergence of life farther inwards, to 4.416–4.52 Ga.

## 3. Habitability of the Earth

### 3.1 Radiative Cooling of the Magma Ocean

The energy deposited by the Moon-forming impact melted the proto-Earth's mantle and caused a global magma ocean (Jacobson et al. 2014; Touboul et al. 2007). Hydrodynamic (i.e. fluid flow) simulations of Moon-forming impacts on the nearly fully-formed Earth show post-impact mantle temperatures in excess of 7000 K (Canup 2004). Any solid mantle that survived the impact would have likely sunk down to the hot core and then melted (Zahnle et al. 2007). At this stage, the Earth had accreted the majority of its material and began to cool. The atmosphere of the newly-melted Earth, composed of mostly rock vapour and silicate clouds, would have lasted through cooling for ~1000 years (Zahnle et al. 2007). During this time, silicates would condense and rain down onto the Hadean Earth's surface at a rate of a meter per day (Zahnle et al. 2007). As Earth's surface continued to cool, large temperature differences were created between both the surface and the interior, and the atmosphere and the surface, which initiated convection cycles (Monteux et al. 2016, Lebrun et al. 2013, Zahnle et al. 2007). Convection would promote cooling by continually bringing mantle gases to the surface where they can then rise to the top of the silicate clouds and radiate (Zahnle et al. 2007). This means the magma mantle was likely devoid of gases after cooling. Once the early Earth had cooled enough for the silicate clouds to rain out, the early atmosphere would have been composed of the remaining volatiles that were outgassed during convection cooling. For ~0.02–100 Myr this hot atmosphere would have lived above a deep magma ocean (Monteux et al. 2016, Lebrun et al. 2013, Zahnle et al. 2007). Thermal blanketing by the hot atmosphere and the surface temperature of the magma ocean would have determined the rate at which the Earth radiatively cooled. The cooler surface and thicker the atmosphere, the more effective the thermal blanket. To understand this, imagine a hot surface heating up the atmosphere, causing it to puff out. By increasing the volume of the atmosphere, the atmospheric density decreases. Only the top of the atmosphere—where pressures are less than a few tenths of a bar—radiates into space (Kasting 1988), thus by expanding the volume of the top of the atmosphere, more radiation can escape. As the surface cools, the atmosphere cools and contracts, decreasing the volume of the top of the atmosphere, allowing the thermal blanket to trap in more radiation. When the surface of the Earth dropped to some temperature below ~1800 K, a constant, inefficient rate of cooling began (Lebrun et al. 2013, Zahnle et al. 2007).

Tidal heating by the Moon would have also played a role in maintaining the magma ocean (Lebrun et al. 2013, Zahnle et al. 2007). Tidal heating is the result of the different strengths of the Moon's gravitational force on the near vs. far sides of the Earth. These forces are large for two bodies orbiting close to one another. For example, the Moon orbits the Earth at a distance of approximately 100 Moon diameters, and the highest tides on Earth reach up to about 12.2 meters. Tidal forces can cause a spherical body to stretch into an oblate spheroid, which compresses and re-stretches to adjust for rotation and orbit. This stretching and compressing causes friction between solids in the body, which generates heat and allows the solids to melt. The strongest source of friction would be in materials that are solid, but nearly melting (Zahnle et al. 2007). Since the Moon formed, the Earth's tides have been transferring angular momentum from the Earth's spin to the Moon's orbit (Ćuk et al. 2016). Over time this has slowed the Earth's rotation, and expanded of the Moon's orbit (Ćuk et al. 2016), which means during the Hadean, tidal forces were stronger and Earth's tides fluctuated more rapidly (Lathe 2004). For about the first 1.4 million years after Moon formation, tidal forces were strong enough to aid in prolonging the Earth's magma ocean (Zahnle et al. 2007). During this time, tidal heating would have generated about the same to 1 order of magnitude less heat than absorbed sunlight, and 2–3 orders of magnitude more heat than radioisotopes (Zahnle et al. 2007).

Calculations by Monteux et al. (2016), Lebrun et al. (2013), and Zahnle et al. (2007) have shown that the Hadean Earth would have maintained a magma ocean, and remained above the boiling point of water for ~0.02–100 Myr. These calculations consider a variety of effects, including convective heat transport, radiative cooling, absorbed sunlight, water's solubility in liquid basalt, and tidal heating. Once the magma ocean solidified, water condensed and rained down to the solid surface at ~1 m/year (Zahnle et al. 2007) forming Earth's oceans in just a few thousand years. The outer boundary for the emergence of life can therefore be moved inwards, to when the Earth was cool enough to maintain liquid water on a solid crust. This would be between 4.316 and 4.52 Ga if we consider the range in Moon formation ages and take no preference to whether cooling was efficient (20 kyr) or inefficient (100 Myr).

## 3.2 Evidence of a Continental Crust

Zircons ($ZrSiO_4$) are minerals which crystallize from melts, and, due to their durability, represent the oldest preserved terrestrial material. Because zircons have U-Pb ages as old as 4.404 ± 0.008 Ga (Wilde et al. 2001), they can be used to glean insight about the Hadean environment. Samples of 4.01- to 4.37-Gyr-old zircons, representing pieces of the Hadean crust, have been isotopically compared with the bulk silicate Earth (BSE) in attempts to estimate the time at which the Earth's crust solidified (Harrison et al. 2005). The main idea behind this technique is that the formation of a continental crust temporarily halts elemental differentiation, freezing in the ratio of hafnium (Hf) isotopes in the solid crust (and zircons) at that time (Amelin et al. 1999; Harrison et al. 2005). However, since the continental crust is continuously remelted and recycled into the mantle, elemental differentiation continues at a slower rate even after crust has formed, leading to an increasingly depleted mantle. For this reason, we would expect that more recent zircons, formed in the upper mantle, would be more depleted in $^{176}Hf/^{177}Hf$ than zircons that formed from an older, less depleted mantle.

One difficulty with this approach is that the Hf isotope composition varies due to the radioactive decay of $^{176}$Lu to $^{176}$Hf. Zircons are useful in this matter since they incorporate a relatively minute amount of Lu, therefore the effect of radiogenic decay on zircon Hf abundances can be considered negligible (Amelin et al. 1999). By characterizing the deviation of $^{176}$Hf/$^{177}$Hf in zircons of various ages to that of the BSE, it is possible to determine whether the $^{176}$Hf/$^{177}$Hf values follow a trend that is consistent with zircon formation from a slowly depleting mantle reservoir. Large negative $^{176}$Hf/$^{177}$Hf deviations increasing in time are consistent with the formation of a continental crust. Large positive $^{176}$Hf/$^{177}$Hf deviations on the other hand, suggest the zircons were derived from reservoirs with high Lu/Hf ratios, which upon $^{176}$Lu decay to $^{176}$Hf, enrich any forming zircons in Hf. Harrison et al. (2005) measured large positive and negative $^{176}$Hf/$^{177}$Hf deviations from the BSE in zircons with ages as old as 4.37 Ga. However, temporal extrapolations of these results that align with today's rock record suggests that a major differentiation event took place between 4.4 and 4.5 Ga followed by continental crust growth and recycling. Since the Moon-forming impact that melted the Earth's surface likely occurred between 4.416–4.52 Ga, the zircon isotopic analysis performed by Harrison et al. (2005) is consistent with the ~0.02–100 Myr magma ocean solidification times calculated by Monteux et al. (2016), Lebrun et al. (2013), and Zahnle et al. (2007).

*3.3 Snowball Earth?*

As the Earth's mantle solidified and subduction of crustal plates began to occur, $CO_2$ would have been removed from the atmosphere due to its high reactivity with silicates (Sleep & Zahnle 2001), and subducted into the Earth's crust. If $CO_2$ subduction occurred early, it could have continued to remove $CO_2$ from the atmosphere until the partial pressure was well below 1 bar, which is about the pressure required to maintain a clement climate on Earth. However, there is still no consensus as to when these plate-tectonic processes began (Maruyama et al. 2017, Ernst et al. 2016, Shirey et al. 2008). Ernst et al. (2016) argue, that the Earth's surface was covered by crustal platelets by 4.4 Ga, and that episodic deep mantle- and plume-driven subduction began as early as 4.4–4.0 Ga. They reason that the terrestrial heat budget requires overturn of platelets during that time, since conduction is much less efficient than convection-advection at transferring heat. Maruyama et al. (2017) argue that an intense period of meteoritic bombardment at 4.37–4.20 Ga initiated plate tectonics on Earth by delivering water (a lubricant) and thickening the crust with lower viscosity material. Finally, Shirey et al. (2008) argue that the drastic change in trace element compositions in rocks starting at ~3.9 Ga suggests the initiation of subduction occurred around this time. Consequently, because there is no guarantee that tectonic processes existed during the Hadean, the following discussion about a "Snowball Earth" is purely speculative.

Given that $CO_2$ subduction led to a Hadean atmosphere much less than 1 bar, the Earth would not benefit from solar radiative heat retention due to the greenhouse effect. Moreover, proceeding the solidification of the magma ocean, geothermal heat was likely insignificant to the Earth's climate (Zahnle et al. 2007). Finally, the Sun would have been only ~70–75% as luminous as it is today, leading to an effective Earth surface temperature of < 230 K (Sagan & Mullen 1972, Zahnle et al. 2007, Feulner 2012). As a result, the Earths' oceans would have frozen over and the Earth would be in a perpetual winter until degassing of $CO_2$ from the crust built up pressure in the atmosphere. If $CO_2$ subduction was a quick process, the vastly changing surface conditions may have disrupted the processes of initiating life. If life began via some kind of RNA World, as seems likely (Neveu *et al.* 2013, Higgs & Lehman, 2015) such a frozen environment may have been advantageous. In fact, Attwater *et al.* (2013) experimentally formed RNA polymerase ribozymes in ice, which were capable of catalyzing templated RNA synthesis at temperatures as low as -19 ˚C. However, if the temperature of the snowball Earth was much colder than -19 ˚C, large molecules embedded in the ice would cease to move, and the RNA world would be halted or slowed.

The snowball Earth could have undergone periodic, or permanent melting due to large impactors. Sleep (2010) estimates that it takes an impactor larger than ~300 km to boil part of the ocean, and that a few objects of this size probably impacted the Hadean Earth after the Moon-forming impact. However, this prediction is based on the statistics of small numbers, and it's also possible there were zero impactors of this size (Sleep 2010). Slightly smaller impactors, similar to the largest lunar impactor (~200 km), would have been more likely to impact the early Earth, and would have also warmed its surface.

Large impactors can also contribute to the loss of an atmosphere through atmospheric stripping (Quintana et al. 2016). However, N-body simulations show that it takes energies comparable to the Moon-forming impact to strip half an Earth atmosphere (Quintana et al. 2016). Nothing bigger than ~1/12 the size of Theia impacted the Earth after the Moon-forming impact (Sleep 2010), therefore subduction may have been a more important driver of atmospheric loss leading to a potential snowball Earth.

The main evidence against the occurrence of a snowball Earth is in the zircon record. 4.0- to 4.35-Gyr-old zircon formation temperatures (~644–801 ˚C) suggest at least a portion of the Earth maintained above freezing temperatures during this time (Watson & Harrison 2005). Moreover, 3.91–4.40 Gyr-old zircons enriched in $^{18}O$ would likely require a high level of water-rock interaction, possibly resulting from the presence of Earth oceans at 4.4–3.9 Ga (Mojzsis et al. 2001, Peck et al. 2001, Wilde et al. 2001, Valley et al. 2002, Cavosie et al. 2005). However, the possibility remains that these zircons formed in a sub-surface ocean below an icy crust. Evidence of surface water on the Hadean Earth will be discussed further in *Evidence of Surface Water*.

### 3.4 Evidence of Surface Water

There are three stable isotopes of oxygen: $^{16}O$ is the most common, followed by $^{18}O$ and then $^{17}O$. Zircons crystals preserve the $^{18}O/^{16}O$ ratio inherited from its magmatic source. $\delta^{18}O$ is a measure of the $^{18}O/^{16}O$ ratio with respect to the standard (see Equation 1 below for a mathematical description). If a zircon formed from magma that was in equilibrium with the primitive mantle, it will have a very narrow range of $^{18}O/^{16}O$ ratios. Specifically, this range is $\delta^{18}O = 5.3 \pm 0.3$ ‰ (Cavosie et al. 2005, Valley et al. 2005). Low temperature liquid water-rock interactions lead to the preferential partitioning of $^{18}O$ into minerals, because the bonds formed by $^{18}O$ in minerals are stronger than those formed by $^{16}O$ (Bindeman 2008). Therefore, if a zircon formed from magma whose parent rock had prolonged contact with low temperature $H_2O$, the water-rock interactions will have enriched the parent rock in heavy oxygen (Cavosie et al. 2005, Valley et al. 2005, Mojzsis et al. 2001). It is possible that closed-system crystallization processes can enrich the $\delta^{18}O$ of melts by a maximum of ~1 ‰, therefore only $\delta^{18}O$ zircon values higher than 6.5 ‰ are interpreted as evidence of liquid water interacting with a solid crust (Valley et al. 2005).

$$\delta^{18}\text{O} = \left( \frac{\left( \frac{^{18}O}{^{16}O} \right)_{sample}}{\left( \frac{^{18}O}{^{16}O} \right)_{standard}} - 1 \right) \times 1000‰ \tag{1}$$

$\delta^{18}$O values greater than 6.5 ‰ have been measured in 3.91–4.40 Ga zircons, and have been interpreted as evidence for surface water on the Hadean Earth as early as 4.4 Ga (Mojzsis et al. 2001, Peck et al. 2001, Wilde et al. 2001, Valley et al. 2002, Cavosie et al. 2005). Liquid water was therefore likely present consistently throughout the Hadean eon.

Evidence of liquid surface water from the zircon record is consistent with the habitability boundary constrained thus far (4.316–4.52 Ga). Because there are no discovered zircons with ages older than 4.4 Ga, this technique cannot determine whether surface water was present on the Earth before that time.

### 3.5 The Late Heavy Bombardment

A short and intense bombardment of asteroids and comets may have occurred on the early Earth at around 3.9 Ga. This hypothesis is termed the late heavy bombardment (LHB), or the late lunar cataclysm. The occurrence of the LHB, and therefore its role in the origin of life, remains contentious. The strongest evidence for the LHB came from data taken by the Apollo program. $^{40}$Ar-$^{39}$Ar isotopes have been used to date the lunar basins, which were formed by large asteroid impacts at 3.85 Ga (Imbrium), 3.89 Ga (Serenitatis), 3.91 Ga (Crisium), and 3.92 Ga (Nectaris) (Ryder et al. 2000). These basin dates are supportive of a cluster of 100 km-sized asteroid impacts around 3.9 Ga. Furthermore, because the probability of asteroid impacts decreases with increasing impactor size, the Moon was likely also struck by hundreds of 10 km-sized asteroids at that time (Wilhelms et al. 1987). Given that the Earth has a gravitational cross-section 20 times the size of the Moon, it was likely bombarded by 20 times as many asteroids, some which would have been much bigger than the biggest to have impacted the Moon.

However, this interpretation is controversial, as these basins could have also formed from asteroid impacts that made up the tail end of a more sustained, declining bombardment (Zahnle et al. 2007). In fact, results from the Lunar Reconnaissance Orbiter suggest the impact melts collected from the Serenitatis basin may be ejecta from the Imbrium basin-forming impact (Spudis et al. 2011), meaning Serenitatis could be much older than traditionally thought. Furthermore, Boehnke & Harrison (2016) developed a physical model of $^{40}$Ar diffusion in Apollo samples, and found that $^{40}$Ar/$^{39}$Ar basin age histograms tend to show illusory LHB-type episodes under monotonically declining impacts. Therefore, lunar magmatism and Ar overprinting from subsequent impacts may cause us to misinterpret the cluster of lunar basin ages as a single lunar cataclysm.

Evidence for a sustained, declining bombardment dating back to the accretion of the Moon is however lacking. For example, the lack of impact-melted lunar samples dating prior to ~3.92 Ga doesn't support a sustained bombardment (Ryder et al. 2000). This may however be due to the fact that impact glasses would shatter into smaller pieces gradually with time given a sustained lunar bombardment (Zellner & Delano 2015). Lunar meteorites recovered from Antarctica also have ages no older than 3.92 Ga, which supports the concept of a short intense bombardment period (Cohen et al. 2000). There is however another possible explanation, that a bias is introduced in sampling Antarctic meteorites; strong rocks near the surface of the Moon that can survive an impact, be ejected towards the Earth, and survive atmospheric entry, are rare, and only become rarer with age (Zahnle et al. 2007).

Astrophysical simulations also support the occurrence of a LHB. In the Nice model of the formation of the Solar System (Tsiganis et al. 2005, Morbidelli et al. 2005, Gomes et al. 2005), as a result of gravitational interactions with the planetesimal disk, the giant planets begin to migrate inwards or outwards. This planetary migration leads to a 1:2 mean motion resonance between Jupiter and Saturn (i.e. Saturn's period becomes twice that of Jupiter's), which forces the orbits of Neptune and Uranus to destabilize into eccentric ellipses, allowing these planets to cross through the planetesimal disk sitting at least 1–1.5 AU beyond Neptune's orbit. The crossing of the ice giants into the planetesimal disk gravitationally disperses the planetesimals, causing many of them to be sent on Earth- and Moon-impacting trajectories. Following this LHB, which occurs approximately 450–850 million years after the formation of the Solar System (Gomes et al. 2005, Bottke et al. 2012), the giant planets settle into separations, eccentricities, and inclinations similar to their observational values (Tsiganis et al. 2005, Morbidelli et al. 2007). The Nice model also correctly explains the orbital distribution and total mass of the Trojan asteroids (Morbidelli et al. 2005).

In Table 2 we summarize the main pieces of evidence that generally support either the LHB or a sustained declining bombardment on the early Earth. All things considered, there is insufficient evidence to reject the occurrence of a LHB, however more lunar research is required before a consensus in the field can be reached. Four possible models for the late lunar bombardment are illustrated in Figure 2 which are calibrated to the crater counts and surface ages at the Apollo landing sites (figure reproduced from Zahnle et al. 2007).

The bombardment of the early Earth with asteroids and comets could have had both positive and negative effects with respect to the emergence of life. If $CO_2$ subduction was efficient during the Hadean eon, asteroidal and cometary impacts on the early Earth could have helped maintain the atmospheric pressure necessary for a clement temperature by releasing $CO_2$ into the atmosphere. However, weathering of the ejecta from such impacts also removes $CO_2$ from the atmosphere (Sleep & Zahnle 2001), possibly much more efficiently than $CO_2$ is released by impacts (Charnay et al. 2017). Impacts could have also delivered or even produced the essential organics required for the emergence of life, e.g. nucleobases, amino acids, and carboxylic acids (Chyba & Sagan 1992, Callahan et al. 2011, Burton et al. 2012, Pizzarello et al. 2012, Martins et al. 2013, Cobb & Pudritz 2014, Pearce & Pudritz 2015). On the other hand, asteroidal and cometary impacts could have also hindered the progress of evolving pre-life molecules or destroyed existing life (Zellner 2017).

Abramov et al. (2013) developed a thermal model for global bombardments to study how significant Earth's crustal melting would have been due to the occurrence of the LHB. Their model, which considered shock heating, impact melt generation, uplift, and ejecta heating, predicted that about 5–10 % of the Earth's surface area would have been covered by >1 km deep impact melt sheets after the LHB. This suggests that living organisms could have potentially survived the event if they inhabited a portion of the surface that remained relatively unscathed. But what if the LHB was just the tail-end of a sustained, declining bombardment? Given this scenario, a higher proportion of the crust would have melted from 4.5–3.9 Ga than would have during a LHB, however it is unclear how much greater this proportion would be. The existence of zircons that formed from melts with ages from 3.91–4.40 Ga confirms that the Earth's crust experienced periodic melting throughout this period, but more importantly, since re-melting of the crust would have likely age-reset the zircons at that location (Abramov et al. 2013), at least some crust likely didn't re-melt from 4.4–3.9 Ga.

**Table 2.** Summary of the main pieces of evidence generally in favour of either the LHB or a sustained declining bombardment on the early Earth.

| Evidence | More in favour of | Source |
|---|---|---|
| Cluster of lunar basins around 3.9 Ga | Late Heavy Bombardment (LHB) | Ryder *et al.* (2000) |
| Lack of lunar meteorites older than 3.92 Ga | LHB | Cohen *et al.* (2000) |
| Planetary dynamics simulations (Nice Model) | LHB | Gomes *et al.* (2005) |
| Potential incorrect dating of Serenitatis basin | Sustained declining bombardment | Spudis *et al.* (2011) |
| Physical model of $^{40}$Ar diffusion in Apollo lunar samples | Sustained declining bombardment | Boehnke & Harrison (2016) |

If there was a LHB and it completely sterilized the planet of any evolving life and pre-life molecules (e.g. nucleotide monomers, dimers, trimers, etc.), the habitability boundary of the emergence of life is ~3.9 Ga. Otherwise, if asteroid impacts at or before 3.9 Ga had no negative bearing on the emergence of life, the outer boundary remains at 4.316–4.52 Ga, corresponding to the uncertainty in the date by which the Earth's magma ocean cooled and a hydrosphere formed.

## 4. The Fossil Record

*4.1 Microfossils*

It has been said that the only true consensus for a date by which life definitely existed on Earth is in the bacterial fossils found within the Gunflint Formation of Ontario, dated at 1.9 Ga (Moorbath 2005, Wacey *et al.* 2011). Brasier *et al.* (2015), a prominent challenger of the oldest microfossils reaffirmed this, calling the Gunflint chert a benchmark for the analysis of early fossil preservation. Two key components of the Gunflint microfossil consensus are that (a) these microfossils have 3-dimensional round-walled compartments, a.k.a., cell lumina, and (b) the walls are composed of carbonaceous (kerogenous) matter (Schopf 2007, Igisu et al. 2009). These two traits above others, have become critical evidence in establishing the authenticity of microfossils (Schopf & Kudryavtsev 2012). However, authentic microfossils are also backed by other more common indicators such as the presence of a variety of specimens ranging from relatively well-preserved to partially or completely degraded (Schopf *et al.* 2010). For a complete list of microfossil indicators, see Wacey (2009).

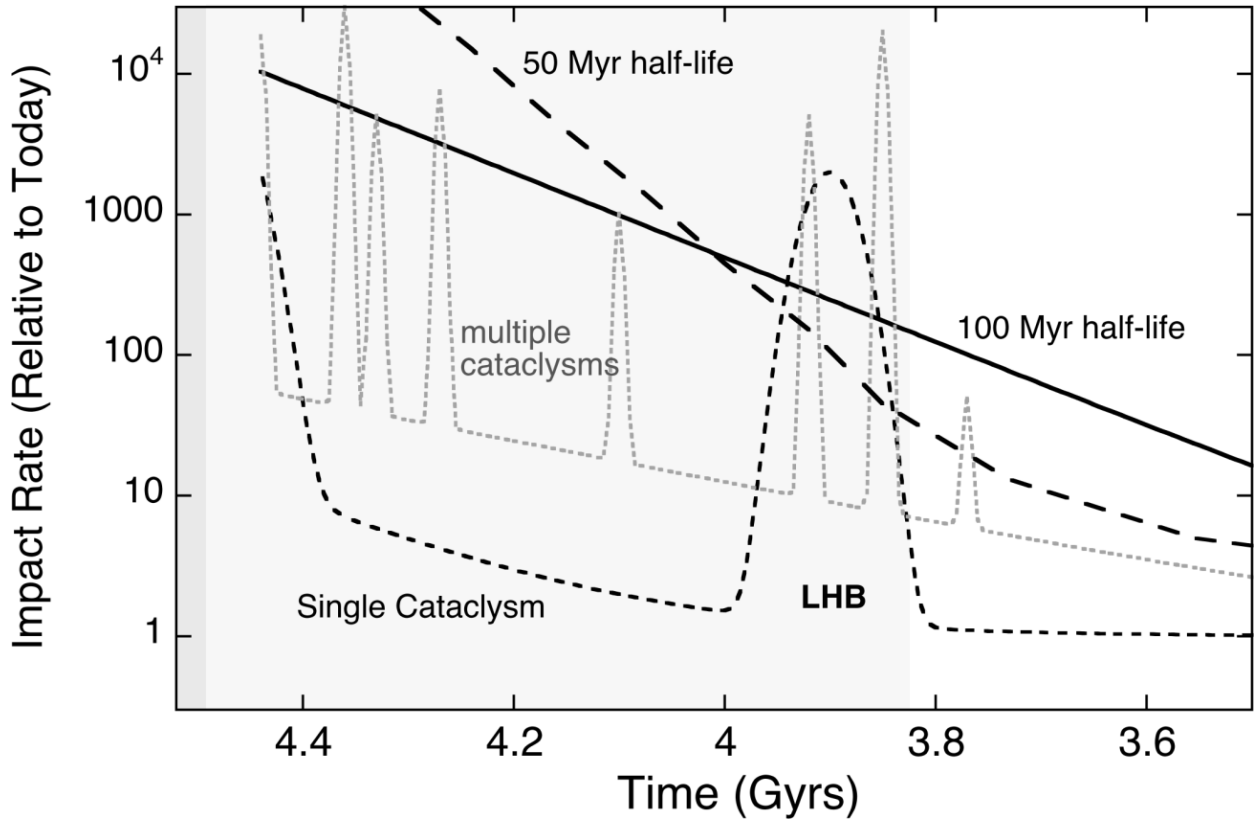

**Figure 2.** Four possible scenarios for the late heavy bombardment, calibrated to crater counts and surface ages at the Apollo landing sites. All scenarios except the 50 Myr half-life model are supported by the available data. Reprinted by permission from Springer Nature: Zahnle *et al.* (2007).

It has been argued by many that microfossils have been discovered within carbonaceous cherts (i.e. the fine-grained silica-rich sedimentary rocks) and sandstones in Western Australia (Schopf & Packer 1987, Schopf 1993, Rasmussen 2000, Ueno *et al.* 2001, Kiyokawa *et al.* 2006, Westall *et al.* 2006a, Glikson *et al.* 2008, Wacey *et al.* 2011, Sugitani *et al.* 2013, Hickman-Lewis *et al.* 2016), within ferruginous sedimentary rocks in Northern Quebec (Dodd *et al.* 2017), and within pillow lavas, siliciclastic deposits (noncarbonate, almost exclusively silica-bearing sedimentary rocks), and cherts in South Africa (Knoll & Barghoorn 1977, Walsh & Lowe 1985, Walsh 1992, Westall *et al.* 2001, Furnes et al. 2004, Westall *et al.* 2006b, Javaux *et al.* 2010) that range in diameter from ~0.1–289 µm. The rocks in the formations hosting these possible microfossils have U-Pb dates of approximately 3.20–3.77 Ga. (For a more complete review of potential microfossils, including those younger than 3.2 Ga, see Schirrmeister *et al.* 2016.) These potential microfossils come in many morphologies: spiral, branched, thread-like, disc, spheroidal, cylindrical, tubular, and filamentary. We present a summary of the authenticity of these potential microfossils in Table 3.

**Table 3:** Summary of ≥ 3.2-Gyr-old potential microfossils and how they fare against the most critical traits for microfossil authenticity.

| Location discovered | Age (Ga) | Cell lumina and carbonaceous cell walls confirmed with electron microscope and mass spectrometer? | Source(s) |
|---|---|---|---|
| Moodies Group, Clutha Formation | 3.2 | Yes | Javaux *et al.* (2010) |
| Pilbara Supergroup, Dixon island Formation | 3.2 | No | Kiyokawa *et al.* (2006) |
| Pilbara Supergroup, Kangaroo Caves Formation | 3.24 | No | Rasmussen (2000) |
| Swaziland Supergroup, Swartkoppie Formation | 3.26 | No | Knoll & Barghoorn (1977) |
| Swaziland Supergroup, Kromberg Formation | 3.4 | No | Walsh (1992), Westall *et al.* (2001), Furnes *et al.* (2004) |
| Pilbara Supergroup, Strelley Pool Formation | 3.43 | No | Sugitani *et al.* (2013) |
| Pilbara Supergroup, Strelley Pool Formation | 3.43 | Yes | Wacey *et al.* (2011) |
| Pilbara Supergroup, Panorama Formation | 3.45 | No | Westall *et al.* (2006a) |
| Swaziland Supergroup, Josefsdal Formation | 3.45 | No | Westall *et al.* (2006b) |
| Swaziland Supergroup, Hoogenoeg Formation | 3.45 | No | Walsh & Lowe (1985), Furnes *et al.* (2004) |
| Pilbara Supergroup, Apex Basalt | 3.46 | No | Schopf & Packer (1987), Schopf (1993), Hickman-Lewis *et al.* (2016) |

| Pilbara Supergroup, Dresser Formation | 3.49 | No | Ueno *et al.* (2001), Glikson et al. (2008) |
|---|---|---|---|
| Nuvvuagittuq supracrustal belt | > 3.77 | No | Dodd *et al.* (2017) |

The difficulty in determining the authenticity of microfossils, is that there are abiotic processes that can lead to similar structures (French & Blake 2016, Brasier et al. 2015, Wacey 2009, Sugitani et al. 2007, García-Ruiz et al. 2003, Brasier et al. 2002). Brasier et al. (2002) reinterpret the potential microfossils discovered in the Apex Basalt of Western Austrailia (Schopf & Packer 1987, Schopf 1993) as being secondary artefacts formed by amorphous graphite reorganized in the form of filaments within multiple generations of hydrothermal chert breccia veins and volcanic glass. These potential microfossils showed similar morphologies to structures within rocks that were hydrothermally altered or completely melted, which questions a biological origin of the possible microfossil morphologies (Brasier et al. 2002). To show this, Brasier et al. (2002) compared these potential microfossils with several similar structures produced in both the clasts (fragments) and matrix of hydrothermally altered chert beds, within glass rims of volcanic shards, and in the associated chalcedony (i.e. milky quartz) matrix of the felsic tuffs (i.e. igneous rocks, rich in elements producing feldspar, fused from smaller grains by heat). These potential microfossil structures (Schopf & Packer 1987, Schopf 1993) are also generally isolated, irregularly distributed, and randomly oriented in several generations of hydrothermally altered formations. This is unlike typical cyanobacteria, in which the trichomes (hair-like outgrowths) of such organisms commonly cluster together in layers, often with a specific orientation relative to rock bedding (Brasier et al. 2002).

Furthermore, García-Ruiz et al. (2003) demonstrated the abiotic formation of noncrystallographic, curved, helical morphologies, similar to these same potential microfossils (Schopf & Packer 1987, Schopf 1993). García-Ruiz et al. (2003) grew filamentous materials out of barium salt in alkaline sodium silicate solutions under alkaline, mildly hydrothermal conditions. These initial materials and conditions, though uncommon on Earth today, are geochemically plausible during Archaean times (García-Ruiz 1998). French & Blake (2016) similarly found a plausible abiotic formation mechanism for microfossils by examining submarine glasses from the western North Atlantic Ocean. By combining petrographic and electron microscopic observations with theoretical models of radiation damage from uranium and thorium decay, French & Blake (2016) found that tubular and granular microfossils can be formed by preferential seawater corrosion of damage trails left by fission fragments.

Given the circumstances, it is difficult to be certain of a biological origin for microfossils without evidence for cell lumina and carbonaceous cell walls. The most recent studies have used state-of-the-art analysis techniques including laser Raman microspectroscopy, secondary ion mass spectrometry (SIMS), and transmission electron microscopy (TEM) to confirm these two traits. The oldest microfossils to meet these criteria are 3.2 (Javaux *et al.* 2010), and 3.4 Gyr old (Wacey *et al.* 2011). In Figure 3, optical photomicrographs of several hollow, tubular microfossils from the 3.4 Gyr-old Strelley Pool Chert are displayed (figure from Wacey *et al.* (2011)). The closely packed tubular microfossils in images (b), (d), and (e) are interpreted by Wacey *et al.* (2011) to be much like the cells of prokaryotes within modern biofilms.

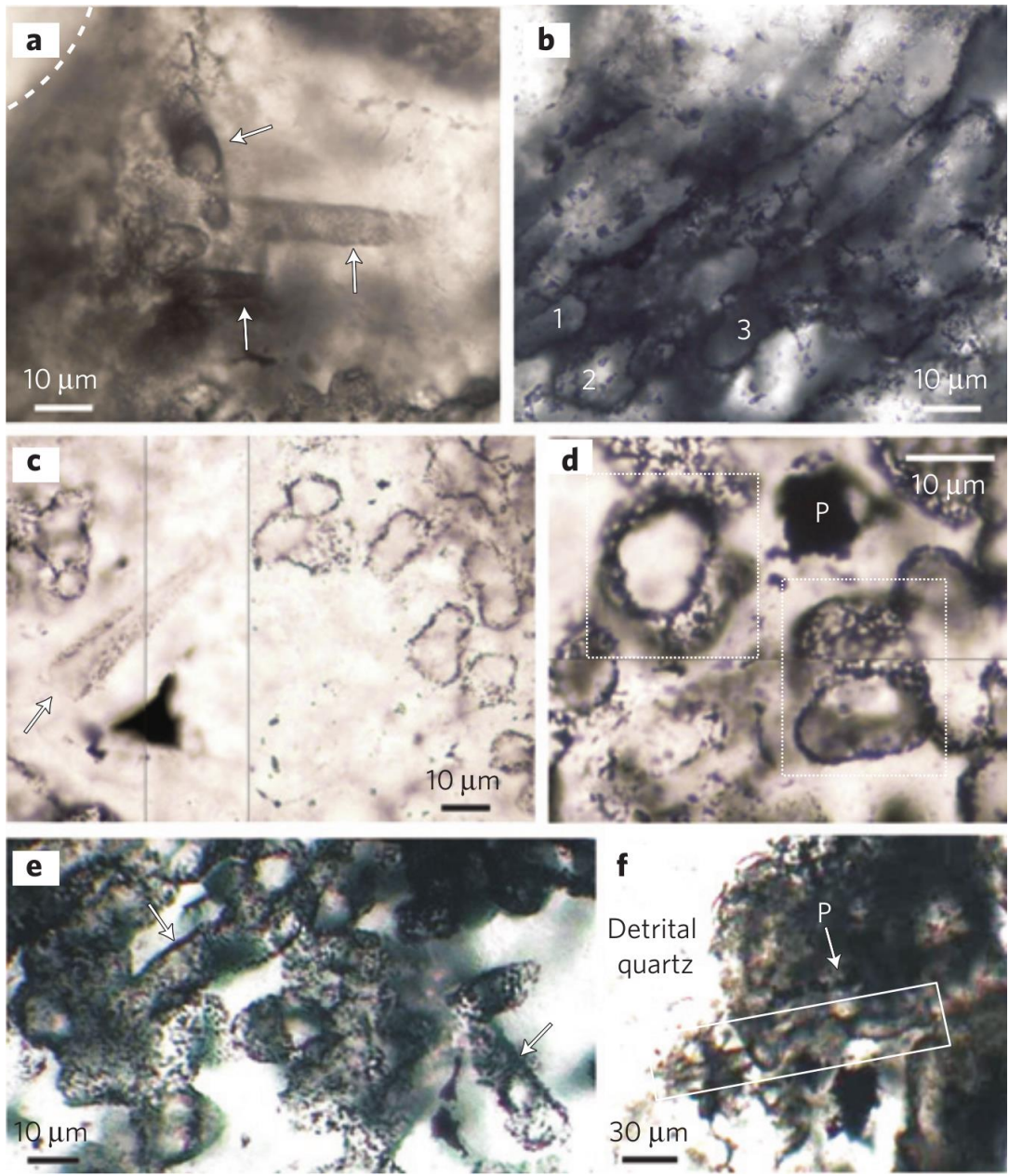

**Figure 3. (a–f)** Hollow, tubular microfossils from the 3.4 Gyr-old Strelley Pool formation (indicated by arrows, boxes, or numbers). A couple highlights are the tubular microfossil cross-sections in frame (d) and the dense patch of tubular microfossils in frame (e). P stands for pyrite. Reprinted by permission from Springer Nature: Wacey *et al.* (2011).

The 3.46 Gyr-old microfossil-like structures in the Apex chert were confirmed by De Gregorio *et al.* (2009) to meet one of the two traits required for establishing authenticity (i.e. carbonaceous cell walls). However, Brasier et al. (2015) examined thin slices of these microfossils with TEM and found at high spatial resolution, instead of cell lumina, the potential microfossils actually have a complex, incoherent spiky morphology, which is evidently formed by filaments of clay crystals coated with iron and carbon. This analysis led them to the conclusion that the multiple clay minerals within the thin slices are entirely compatible with chemistry in high temperature hydrothermal settings.

Lastly, although the recently discovered filamentary and tubular microfossil-like structures in the > 3.77 Gyr-old Nuvvuagittuq belt co-occur with carbonaceous material (Dodd *et al.* 2017), they do not exhibit carbonaceous cell walls and there is no evidence of cell lumina. Therefore these structures do not meet the most critical traits for establishing microfossil authenticity.

In light of all this evidence, in Figure 1, we date the oldest microfossils as 3.4 Ga, which matches the oldest microfossils exhibiting cell lumina and carbonaceous cell walls (Wacey *et al.* 2011). We now discuss the stromatolite macrofossils, which are also useful in constraining the biosignature boundary.

*4.2 Stromatolites*

The word 'stromatolite' in Greek literally means, layered rock. In general, stromatolites are layered, sheet-like, accretionary structures created by or resembling those created by microbial mats of microorganisms such as cyanobacteria. The suite of microbes that form stromatolites live in a biofilm bound by mucus or other adhesives that microbes produce. When photosynthetic microorganisms at the top of the mat get covered in naturally accumulating sedimentary grains and silt, and microbiologically induced carbonate precipitates, they migrate upwards towards the light. This creates a new microbial mat layer, with a calcium carbonate-cemented layer of sedimentary rock and silt left behind (McNamara and Awaramik, 1992, Altermann *et al.* 2006). Over thousands of years, layer upon layer of this combination of sedimentary rock, silt and carbonate builds up. These preserved structures are stromatolites.

The oldest stromatolite-like structures were recently discovered in the 3.7-Gyr-old Isua supracrustal belt in Greenland (Nutman et al. 2016). A photograph of these stromatolites along with the interpretation by Nutman et al. (2016) and a comparison to similar, 2.03-Gyr-old stromatolites are shown in Figure 4. Previous to this discovery, the oldest widely accepted stromatolites were found in the Strelley Pool Chert (Allwood et al. 2006), the contact between the Strelley Pool Chert and Panorama Formation (Hoffman 1999), and the Dresser Formation (Noffke et al. 2013) in Western Australia. Rocks from these locations are U-Pb dated at ca. 3.43 Ga, 3.45 Ga and 3.48 Ga, respectively. Most recently, Djokic et al. (2017) also discovered exceptionally well-preserved stromatolites in the 3.48 Gyr-old Dresser Formation. Nutman et al. (2016) deduce that the 3.7-Gyr-old stromatolite forms match some of those discovered in the 3.43-Gyr-old Strelley Pool chert and the 2.03-Gyr-old Wooly Dolomite, implying a biological origin. Since the oldest stromatolites don't contain any fossil microbes, many assume that they are of biological origin due to morphological comparisons with modern, biological stromatolites (Grotzinger & Rothman 1996). Some key morphological indicators are wrinkly sequences of small-scale fine layers in cone, mat, peak and dome shapes and microbe-bound ripples in carbonate sand (Nutman et al. 2016; Van Kranendonk 2011; Allwood et al. 2006).

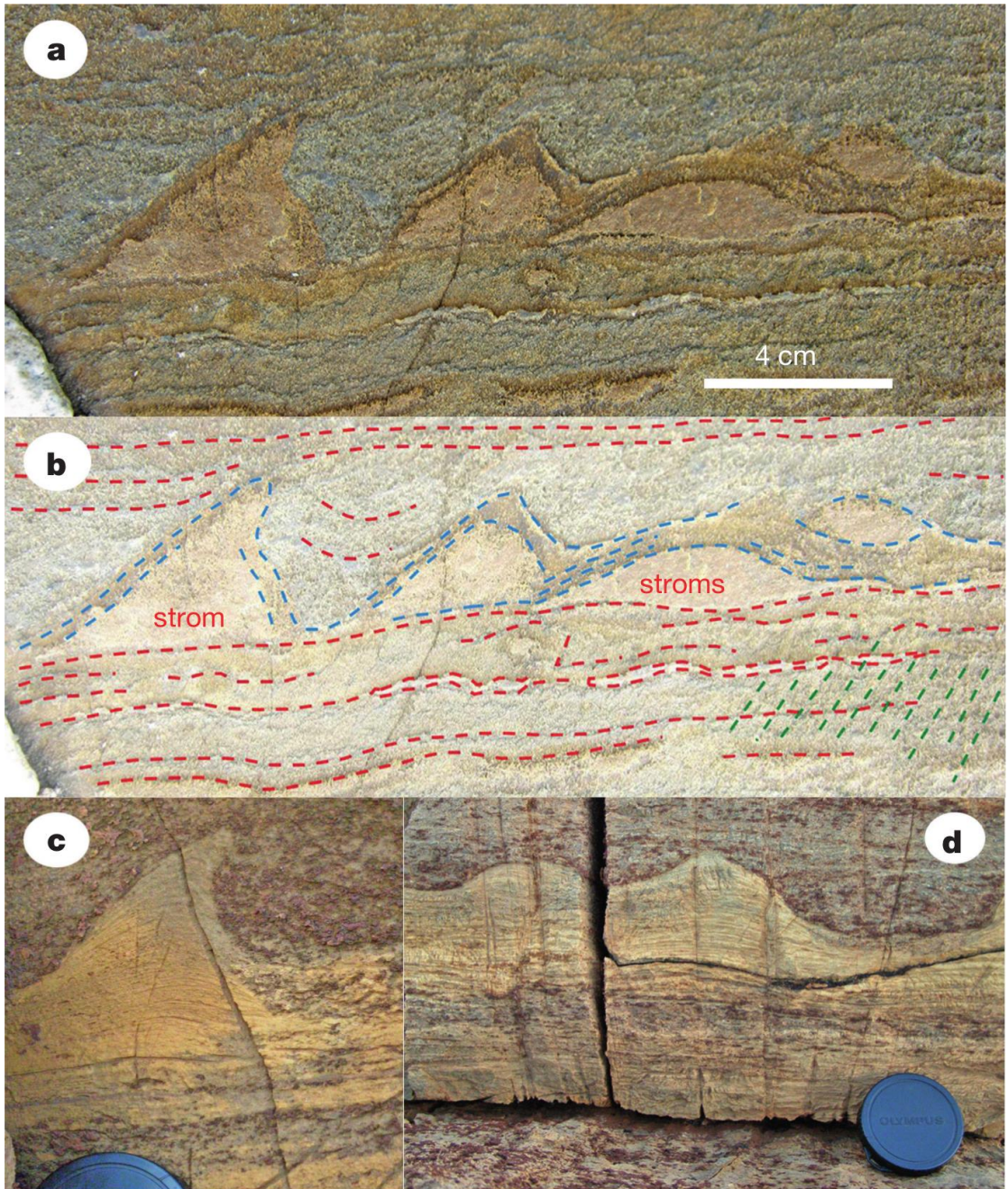

**Figure 4. (a)** Stromatolites in an exposed outcrop of 3.7-Gyr-old metacarbonate rocks in the Isua supracrustal belt in southwest Greenland. **(b)** Interpretation of (a), with both an isolated stromatolite (strom) and a conglomerate of stromatolites (stroms). **(c)** Asymmetrical, peak-shaped stromatolite and **(d)** linked, dome-shaped stromatolites from the ~2-Gyr-old Wooly Dolomite carbonate platform in Western Austrailia. Reprinted by permission from Springer Nature: Nutman *et al.* (2016)

Just as with microfossils, there has also been some disagreement whether stromatolite rock formations are of biological origin. For example, Grotzinger & Rothman (1996) deduce that these surfaces can be grown from chemical precipitation, diffusive rearrangement of suspended sediment, and uncorrelated random noise. Others defend that the distinct morphological attributes of stromatolites are inconsistent with purely mechanical deposition (Nutman et al. 2016, Allwood et al. 2006). For example, unlike crusts formed by mechanical deposition, the peaks of the 3.7 Ga stromatolite structures are inclined and asymmetrical, both dome- and peak-shaped structures are observed together, and there are no irregular projections normal to the growth surface (see Figure 4).

Non-morphological factors also indicate a biological origin of the oldest stromatolites. For example, there is a great diversity of stromatolites in the same location that all resemble known microbial mats, and none resemble any known structure formed abiotically (Allwood et al. 2006). This is consistent with the ecologically controlled growth of an Archean microbial reef (Allwood et al. 2006). These stromatolites also exist in a location where the shoreline moved inwards and deposited carbonates: a location where microbial mats exist today (Peterffy et al. 2016). There is also an absence of ancient stromatolites in deep waters, where photosynthetic organisms wouldn't have had enough light to exist in microbial mats (Allwood et al. 2006). Altogether, the evidence that stromatolites are true biological fossils seems fairly strong.

With all of this evidence in hand, we are confident in pushing the biosignature boundary for the emergence of life outwards to the date of the oldest stromatolites, at 3.7 Ga.

## 5. Isotopic Signatures

### 5.1 Signatures for the Presence of Oxygen

Observations of sulfur and molybdenum isotopes and transition metals in ancient sedimentary rocks indicate that there was a so-called great oxidation event (GOE) around 2.4–2.3 Ga. The GOE is widely believed to have been produced by photosynthetic cyanobacteria (Planavsky et al. 2014; Buick 2008; Grula 2010; Sessions et al. 2009). Photosynthetic organisms were producing oxygen prior to the GOE, however it is argued that at 2.4–2.3 Ga, the Earth's surface became oxidized to the point where $O_2$ became more stable in the atmosphere than competing reduced gases (Zahnle et al. 2013). UV photodissociation of water vapour can produce $O_2$ abiotically, however unless one of the products of such a reaction is immediately removed, they will recombine to form water vapour (van Andel 1994). Furthermore, water vapour would have been present in the atmosphere throughout the Hadean and Archean eons (see Section 3), and there is no known reason for the $O_2$ produced via this mechanism to start building up around 2.4–2.3 Ga. This argument can also be used to explain why the photodissociation of atmospheric $CO_2$ (Lu et al. 2014) is unlikely to have caused the GOE. Hence, the existence of $O_2$ in the early atmosphere implies a rapid rate of creation of $O_2$ by life that keeps the atmosphere out of equilibrium.

The fluctuations in atmospheric oxygen over geological time scales can be tracked by fluctuations in sulfur and molybdenum isotopes (Kaufmann et al. 2007; Planavsky et al. 2014) and enrichment of transition metals (e.g. chromium, molybdenum, rhenium) in the geologic record (Anbar et al. 2007). Sulfur isotopes are said to undergo mass independent fractionation (MIF) under anoxic conditions. However the term "MIF" can be misleading, as fractionation of sulfur isotopes is always dependent on the mass of the isotopes. Literally, MIF just means the isotopic signatures don't obey the standard mass-dependent relationship (i.e. $\delta^{33}S \sim 0.515\delta^{34}S$; $\delta^{36}S \sim 1.91\delta^{34}S$) (Farquhar et al. 2000, Pavlov & Kasting 2002). The reason for this deviation from the standard isotopic relationship in anoxic environments, is that when sulfur rains out of the atmosphere or is deposited onto the surface, it does so in a variety of different oxidation states—leaving a variety of isotopic signatures in the sediments into which sulfur gets incorporated. In an oxygen atmosphere, even one as low as $10^{-5}$ times the present atmospheric level of $O_2$, the clear majority of sulfur gets oxidized to $H_2SO_4$, which homogenizes the sulfur isotopes to match the standard mass-dependent relationship (Pavlov & Kasting 2002). Therefore measuring such isotopes in old bioelemental sediments can be used to infer the oxygenation history of ancient seawater (Pufahl & Hiatt 2012). An accumulation of redox-sensitive transition metals in sedimentary rocks also signifies a necessity of oxygen in the early Earth seawater that deposited such rocks (Anbar et al. 2007). For example, molybdenum, which currently exists in bodies of water as the unreactive molybdate ion ($MoO_4^{2-}$), accumulates via oxidative weathering of molybdenum-bearing sulfide minerals in crustal rocks. Under anoxic conditions, molybdenum would instead be retained by these sulfide minerals and therefore would not accumulate in the oceans and be deposited into sediments (Anbar et al. 2007).

MIF disappears completely from the geologic record before 2.4–2.3 Ga, providing the tightest constraint on the GOE (Sessions et al. 2009). A shift in sulfur isotopes (Kaufmann et al. 2007) and an enrichment of transition metals (Anbar et al. 2007) in 2.5-Gyr-old samples of shale from Mount MacRae in Western Australia indicates a presence of atmospheric oxygen even before the GOE. Moreover, Planavsky et al. (2014) measured a large fractionation of molybdenum isotopes in 2.95-Gyr-old rocks from the Sinqeni Formation in South Africa. However in this case, molybdenum is used as a proxy for manganese (II), which requires free dissolved $O_2$ to be oxidized. Planavsky et al. (2014) inferred from these results that oxygen produced via photosynthesis began to accumulate in shallow marine settings at 2.95 Ga.

This is the kind of environment where we would expect stromatolites to grow, and as we saw in *Stromatolites*, evidence for stromatolites goes back to 3.7 Ga, about three-quarters of a billion years before the $O_2$ signature is seen. These dates do not seem inconsistent to us. Stromatolites are presumed to be formed by photosynthetic bacteria, and modern stromatolites contain cyanobacteria that carry out oxygenic photosynthesis. However, there are several other groups of bacteria that carry out non-oxygenic photosynthesis (see Section 6 below). The earliest stromatolites might have been formed by bacteria such as these, or it may simply be that oxygenic photosynthesizers were present but oxygen sinks (e.g. organic matter, iron) near the Earth's surface at this time efficiently removed atmospheric $O_2$ (Zahnle *et al*. 2013). Thus, the evidence from $O_2$-dependent isotopic signatures seems to fit with that from the fossil record, but it does not push the biosignature boundary further back from what is known from stromatolites. In contrast, carbon isotope biosignatures extend further back in time and are currently a key piece of evidence that constrains the biosignature boundary, as we now discuss.

*5.2 Carbon Isotope Biosignatures*

Almost all rocks older than 3.6 Gyr, deposited just above the basement of the crust, underwent high-grade metamorphism (i.e. changed in mineral assemblage and structure due to high temperatures and pressures). Therefore fossils from before 3.6 Ga are very unlikely to be preserved (which makes the 3.7-Gyr-old stromatolite discovery both remarkable and potentially controversial). This means typically, evidence for life beyond 3.6 Ga must include a geochemical component (Rosing 1999).

One isotope commonly used to infer evidence of biological activity is $^{13}$C. $^{13}$C is a stable isotope of carbon, and makes up approximately 1.1% of the natural carbon on Earth. The rest of the carbon is in the form of $^{12}$C, with only trace amounts existing of the short-lived radioactive $^{14}$C.

Two pioneering studies performed by Nier & Gulbransen (1939) and Murphey & Nier (1941) discovered that converting inorganic carbon to biological matter through biochemical reactions leads to an obvious fractionation of $^{12}$C and $^{13}$C. Subsequent works determined that photosynthetic reaction pathways discriminate against $^{13}$C because of its heavier mass. In general, isotopically light molecules are more mobile and tend to have greater velocities than their heavier counterparts (White & Irvine 1998). As a result, not only does $^{13}$C diffuse slower than $^{12}$C through membranes, but certain enzymes fix lighter carbon faster, e.g. rubulose 1,5-bisphosphate carboxylase and pyruvate dehydrogenase (Subbarao & Johansen 2001, White & Irvine 1998, DeNiro & Epstein 1977). However, not *all* enzymes prefer light carbon, for example, phosphoenolpyruvate carboxylase does not discriminate between $^{12}$C and $^{13}$C (Subbarao & Johansen 2001). Because of discriminatory enzymes, molecules produced by living organisms are preferentially composed of light carbon, while heavy carbon is retained in the surface reservoir (mostly in the form of marine bicarbonate) (Schidlowski 1988). It is unknown when these discriminatory enzymes were incorporated into life, therefore there may be a limit to the earliest organisms that can be detected using carbon isotopes.

The variable $\delta^{13}$C is used to determine how depleted or enriched a sample is in $^{13}$C (similarly to the $\delta^{18}$O variable from Equation 1). $\delta^{13}$C is calculated as

$$\delta^{13}\text{C} = \left( \frac{\left( \frac{^{13}C}{^{12}C} \right)_{sample}}{\left( \frac{^{13}C}{^{12}C} \right)_{standard}} - 1 \right) \times 1000‰ \qquad (2)$$

A comparison of the $\delta^{13}$C values of major groups of plants and autotrophic microbes, and inorganic carbon of the surface environment ($CO_2$, $HCO_3^-$, $CO_3^{2-}$) are displayed in Figure 5 (figure from Schirrmeister et al. 2016). A negative $\delta^{13}$C value corresponds to a sample depleted in $^{13}$C (or enriched in $^{12}$C) with respect to the inorganic, oxidized carbon sources.

Because molecules produced by modern living organisms are preferentially light in carbon, measuring highly negative $\delta^{13}$C values in organic matter within rocks of sedimentary origin is suggestive of biological activity at the U-Pb date at which the rocks are aged. A sedimentary origin is an essential prerequisite for interpreting $\delta^{13}$C signatures to be biologically reduced, because there are abiotic mechanisms for reducing $\delta^{13}$C in rocks that are of non-sedimentary origin, and because a sedimentary origin provides evidence for a liquid water biosphere in a habitable temperature range. This is in contrast to rocks of metasomatic origin, which were formed by hot hydrothermal fluids reacting with rocks originally deposited by metamorphic or igneous fluids. Graphite, for example, is either formed by the metamorphism of carbon-rich sedimentary rocks or from the precipitation of carbon-bearing fluids or melts (Luque et al. 2014). There are several high-temperature abiotic mechanisms that could produce light $\delta^{13}$C in rocks of non-sedimentary origin, including A) disproportionation of ferrous carbonates (van Zuilen et al. 2002), B) decarbonation (the removal of $CO_2$) during metamorphism accompanied by a Rayleigh distillation process or serpentinization (i.e. the addition of water into the crystalline structure of rock minerals) (Fedo & Whitehouse 2002), C) Fischer-Tropsch reactions (McCollom & Seewald 2007), D) diffusive fractionation (Mueller et al. 2014), and E) degassing of basalts (Shilobreeva et al. 2011).

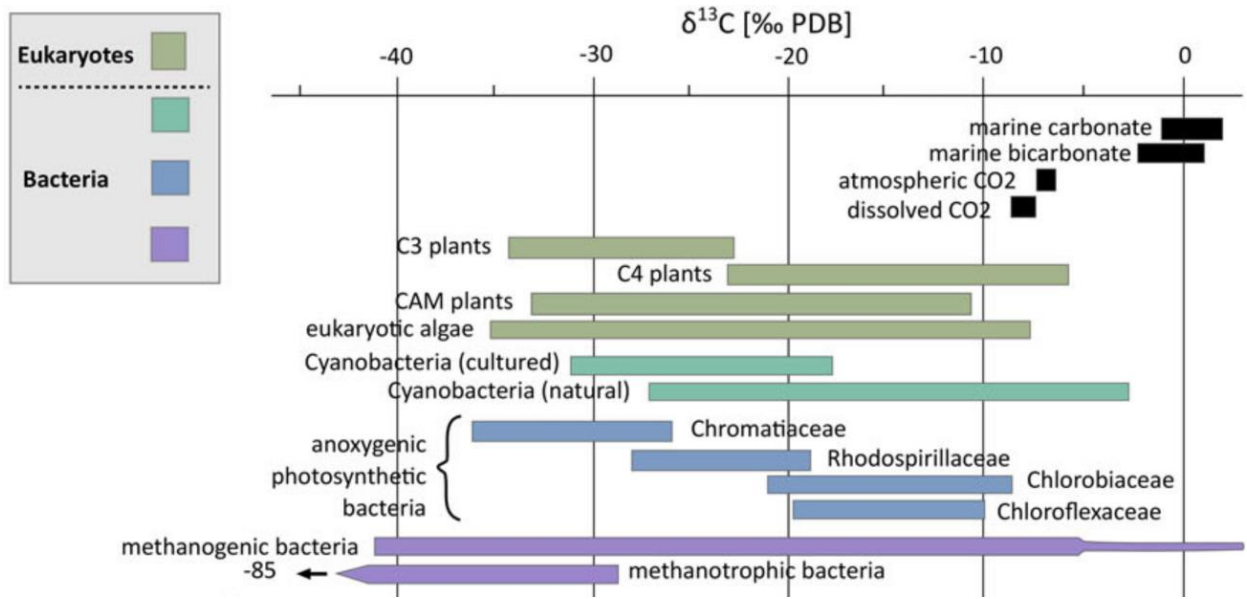

**Figure 5.** Isotopic carbon composition of various major groups of plants, protists, and bacteria compared with those of $CO_2$, $HCO_3^-$, and $CO_3^{2-}$ in the surface environment. Negative $\delta^{13}C$ values correspond to $^{13}C$ depletion relative to the PDB standard. The positive extremes for methanogens were from experiments of cultures that were placed in conditions irrelevant to natural populations. CAM, Crassulacean acid metabolism; PDB, Pee Dee Belemnite. Reprinted by permission from Cambridge University Press: Schirrmeister *et al.* (2016).

By sampling the $\delta^{13}C$ value of graphite globules from sedimentary rocks from the Isua supracrustal belt (ISB) in west Greenland, dated at > 3.7 Ga, Rosing (1999) deduced that they were biologically reduced. The low $\delta^{13}C$ values within the globules (~ -19 ‰), are within the range of reduced C compositions from photosynthetic bacteria (i.e. approximately -10 to -36 ‰, see Figure 5). The same ~3.7-Gyr-old graphite globules were geochemically analyzed by Ohtomo et al. (2014) to confirm whether they were of sedimentary origin, or instead deposited by metamorphic or igneous fluids. With the use of Raman spectroscopy, it was determined that the rocks containing the graphite globules were formed from marine sediments that were depleted in $^{13}C$ at the time of their deposition. Furthermore, through the use of a transmission electron microscope, it was seen that the graphite globules are nanoscale polygonal and tube-like grains, which is unlike the flaky morphology of abiotic graphite in carbonate veins. Even more so, the graphite globules exhibited distorted crystal structures and disordered stacking of graphene, which is consistent with the thermochemical decomposition and pressurization of organics during metamorphism. Therefore the analysis by Ohtomo et al. (2014) supports the claim by Rosing (1999), that the ~3.7-Gyr-old graphite globules are of biological origin.

Grassineau et al. (2006) measured similarly low $\delta^{13}C$ values (~ -18.4 to -14.7 ‰) within samples obtained from a sedimentary outcrop in the > 3.7-Gyr-old ISB, and stated that these samples might record biological activity as early as 3.8 Ga. However due to two high-grade metamorphic events in this outcrop at ~3.74 Ga, metamorphic alteration could have overprinted the $\delta^{13}C$ values within these samples. This means the low $\delta^{13}C$ values in these samples could be a result of high temperature processes rather than the burial of organic matter.

Schidlowski (1988) and Mojzsis et al. (1996) measured mean $\delta^{13}C$ values of approximately -13 to -37 ‰ in deposits of kerogen and grains of apatite from what they interpreted as banded iron formations (BIFs) in the ISB (U-Pb dated to be 3.8 Ga and 3.85 Ga). BIFs are ~4,280–545 Ma layered sedimentary rocks, whose successive layers contain high abundances (20–40%) of iron (Mloszewska et al. 2011, Mloszewska et al. 2013). Both abiotic and biotic sources of the iron precipitate have been suggested, however iron-oxidizing bacteria have been gaining in popularity over the past decade as they can explain iron oxidation in anoxic environments (Smith 2015). The $\delta^{13}C$ purported BIF signatures were initially interpreted to be evidence for early life at 3.8 Ga and 3.85 Ga. However, Fedo & Whitehouse (2002) and van Zuilen et al. (2002) challenged this interpretation by examining the claim that the purported BIF, from which the 3.8- and 3.85-Gyr-old samples of $\delta^{13}C$ were obtained (Schidlowski 1988; Mojzsis et al. 1996), came from sedimentary rocks, chemically precipitated by seawater. They deduced, for various geological, petrological, and geochemical reasons, that the purported BIF rocks were most likely once metasomatic and igneous, and thus could have been reduced in carbon by one of the abiotic mechanisms listed above. Additionally, these previously molten rocks do not provide evidence for a habitable environment, and thus are unlikely to have hosted life.

Papineau et al. (2011) measured $\delta^{13}C$ values from -18.2 to -26.8 ‰ in graphite within the is 3.75- to 4.28-Gyr-old Nuvvuagittuq BIF in northern Canada. These $\delta^{13}C$ signatures may point to a biological origin for this carbon. However, the $^{13}C$-depleted graphite was deposited by hydrothermal or metamorphic fluids after peak metamorphism of the BIF (Papineau et al. 2011, Mloszewska et al. 2013), therefore it could also have been reduced by one of the abiotic mechanisms above. Furthermore, it is uncertain whether this graphite is indigenous to the 3.75- to 4.28-Gyr-old rocks, as it has not undergone the same metamorphic history as the sedimentary host rock (Papineau et al. 2011, Mloszewska et al. 2013).

Finally, preserved graphite encased in 4.1- and 4.25-Gyr-old zircon crystals from Jack Hills, Western Australia have been discovered to have particularly low $\delta^{13}C$ values (Bell et al. 2015, Nemchin et al. 2008). Though both studies could not rule out abiogenic sources, the simplest interpretation may be that the graphite represents organic carbon present at 4.1 and 4.25 Ga. Watson & Harrison (2005) used a thermometer based on titanium content (a "zircon thermometer") to measure the melt temperatures for 54 Jack Hills zircons, and found that they ranged from 644–801 ˚C. At such high formation temperatures, the low $\delta^{13}C$ value in the enclosed graphite could result from any of the five abiotic mechanisms listed above.

With the determination of a likely non-sedimentary origin for the oldest graphite samples described above (those with ages > 3.75 Ga), along with several suggested abiotic mechanisms causing their low $\delta^{13}C$ values, evidence is not strong for the existence of life before 3.7 Ga in Earth's history.

There is also one universal potential abiotic source for low $\delta^{13}C$ values in all old rock samples. This source is carbonaceous meteorites, whose $\delta^{13}C$ contents range from +68 ‰ to -60 ‰ (Bell et al. 2015). If the rock and graphite samples obtained in the ISB are actually fragments of early meteorite impacts, it is possible, given the meteorite fragments are on the low side of their $\delta^{13}C$ range, that the samples are not indicative of biological activity at all. However, this possibility is more likely for rock samples older than ~3.8 Ga, when the possible LHB (or the sustained, declining bombardment) tapered off (see Figure 2). Furthermore, nearly all carbonaceous meteorites contain chondrules (previously molten spherical grains), which are not described to be present near the 3.7-Gyr-old sedimentary rock samples analyzed by Rosing (1999) and Ohtomo et al. (2014).

In Table 4 we summarize biogenicity of the $\delta^{13}C$ signatures in the isotopic studies described above. The general agreement of the sedimentary origin and biological reduction of $^{13}C$ in the ~3.7-Gyr-old graphite globules from Rosing (1999) and Ohtomo et al. (2014) makes us confident in maintaining the biosignature boundary for the emergence of life at 3.7 Ga, as we already concluded from the recently proposed stromatolites in Section 4.2. Although some isotopic evidence may suggest life existed as early as 3.8 Ga (Grassineau et al. 2006), due to the large uncertainty in dates for the sampled outcrop (3.7–3.8 Ga) and the possible metamorphic overprinting of their rock samples at ~3.74 Ga, we are cautious to extend the inner boundary past 3.7 Ga.

**Table 4.** Summary of light $\delta^{13}C$ signatures in > 3.7-Gyr-old rock samples and how they fare against the most critical criterion for biogenicity.

| Location Discovered | Age (Ga) | $\delta^{13}C$ (‰) | Sedimentary Origin? | Source(s) |
|---|---|---|---|---|
| Isua supracrustal belt (ISB) | > 3.7 | -19 | Yes | Rosing (1999), Ohtomo (2014) |
| ISB | > 3.7 | -18.4 to -14.7 | Yes, but potential metamorphic overprinting | Grassineau *et al.* (2006) |
| ISB "Banded-iron formation (BIF)" | 3.8 and 3.85 | -13 to -37 | No | Schidlowski (1988) and Mojzsis *et al.* (1996) |
| Nuvvuagittuq BIF | > 3.75 | -18.2 to -26.8 | No | Papineau et al. (2011) |
| Jack Hills | 4.1 and 4.25 | -5 to -58 | No | Bell *et al.* (2015) and Nemchin *et al.* (2008) |

# 6. Molecular Evolution and Phylogenetics

Another branch of evidence that can provide information about early evolutionary events on Earth is molecular phylogenetics: the use of gene sequences to determine evolutionary relationships among groups of organisms. Important early work on phylogenetics used ribosomal RNA (rRNA), which is one of the few kinds of genes found in all cellular organisms that can be reliably aligned even for the most divergent sequences. Woese and Fox (1977) found that rRNA sequences formed three divergent groups, and thus established the idea that there are three domains of life: Bacteria, Archaea and Eukaryotes. It was also proposed early on, using sequences of pairs of paralogous genes that arose by gene duplication prior to the last common ancestor of the three domains, that the root of the tree of life lies on the bacterial branch (Gogarten *et al.* 1989; Iwabe *et al.* 1989). This implied that the fundamental split in the tree of life is the division between Bacteria and Archaea, with Eukaryotes arising as a sister group to Archaea at a later stage. It has since become apparent that the origin of Eukaryotes involves a fusion between a bacterial and an archaeal ancestor. Ideas of endosymbiosis were proposed very early on (Margulis 1970, Merezhkovsky 1909), and whole-genome analysis in recent years has made it clear that Eukaryotes possess many genes that are of bacterial origin as well as archaeal origin (Lopez-Garcia & Moreira 2015, Koonin & Yutin 2014). Nevertheless, the slowly evolving, highly conserved genes related to the gene-protein translation process show greater similarity between Eukaryotes and Archaea. This is consistent with a scenario in which the root of the translation-related genes lies on the bacterial branch, as originally proposed, and in which other genes were transferred at a later date to the ancestral eukaryote by endosymbiosis. Recent work (Williams *et al.* 2013; Furukawa *et al.* 2017) suggests that eukaryotes arose from within the archaea, not as a sister group. This means that archaea are paraphyletic, and that archaea and eukaryotes should be classed together as a single domain.

The last universal common ancestor of cellular organisms (LUCA) is the earliest point of division among prokaryotic lineages. Studies of early-duplicating paralogues (Gogarten *et al.* 1989, Iwabe *et al.* 1989) suggest that this earliest division lies between bacterial and archeal domain, although an earlier division among lineages of bacteria may also be possible. At what time did this occur? Estimates of dates from phylogenetic trees requires some kind of molecular clock assumption, *i.e.* that mutations occur at a constant rate, so that the number of mutations by which two gene sequences differ is a straightforward function of the time since the species diverged. Geological evidence for species is then used as calibration points to place specific nodes of the tree into geological time. The dates of other points in the tree are then estimated consistently with these calibration points. Sheridan *et al.* (2003) used rRNA sequences to create a phylogenetic tree of major archaeal and bacterial groups, and used a calibration point of 2.65 Ga for the origin of cyanobacteria, based on the detection of 2-methylhopanoids in sedimentary rocks of this age (Summons *et al.* 1999). They estimated a time of 4.29 Ga for the LUCA and 3.46 Ga for both the earliest branch point within Archaea and the earliest branch point within Bacteria. These dates are consistent with other types of evidence that we discussed above. If stromatolites go back to 3.7 Ga, then there must have been a diversity of mat-forming organisms by then. Phylogenetic methods cannot give information prior to the root of the tree; therefore, the 4.29 Ga time estimate for the LUCA is the earliest point accessible from phylogenetics, and the origin of life must be before that. This point is consistent with the estimate of when Earth became habitable after the Moon-forming event (roughly 4.52–4.32 Ga) under the assumption that the LHB did not completely sterilize the Earth at a later date.

It should be realized, however, that there is a large margin for uncertainty in dates derived from molecular clocks, especially when pushing back these methods to the earliest points on the evolutionary tree. The rate of evolution varies substantially between organisms and between genes in the same organism, and these things can be accounted for, to some extent, in estimates of dates (Rambaut and Bromham, 1998; Drummond *et al.* 2006). However, these more sophisticated methods have not yet been attempted for estimating the age of the LUCA, as far as we are aware.

Evolutionary studies of photosynthetic bacteria are particularly relevant to determining the biosignature boundary date because 1) stromatolites are formed by photosynthetic bacteria, 2) low $\delta^{13}$C signatures are taken as an indication of the existence of photosynthetic organisms which use $CO_2$ from the air, and 3) the rise of $O_2$ in the atmosphere is thought to derive from photosynthesis. Oxygenic photosynthesis today is carried out by Cyanobacteria, and by plants and algae which contain chloroplasts derived from Cyanobacteria. The date for the incorporation of plastids into eukaryotic cells is somewhere between 2.1 Ga and 1.2 Ga, depending on which group of fossils is taken to be the earliest eukaryotic algae (McFadden, 2014). This date is substantially later than the origin of Cyanobacteria.

Bacteria performing anoxygenic photosynthesis almost certainly existed before Cyanobacteria. Bacteria capable of anoxygenic photosynthesis include species from the following phyla: Chloroflexi, Chlorobi, Firmicutes, Proteobacteria, Gemmatimonadetes, Acidobacteria and possibly Actinobacteria (Rubrobacter). Gupta (2013) has studied molecular signatures for these different phyla, and concludes that Chloroflexi were the earliest lineage in which photosynthetic ability was fully developed. Chloroflexi are filamentous bacteria, similar to the fossil organisms found in stromatolites. They carry out anoxygenic photosynthesis and could correspond to very early phototrophic microbial communities observed at 3.4 Ga (Tice and Lowe 2006; Olson 2006). Olson (2006) also discusses the geological evidence for the emergence of cyanobacteria carrying out oxygenic photosynthesis by 2.8 Ga, close to the calibration point used by Sheridan *et al.* (2003) discussed above. The emergence of the Chloroflexi group is estimated as 3.1 Ga by Sheridan *et al.* (2003), which is fits reasonably with the scenario above. As we noted in Section 5.1, the dramatic rise in atmospheric oxygen occurring at the GOE, around 2.4–2.3 Ga seems to be somewhat after the first appearance of Cyanobacteria.

Now that many whole genomes have been sequenced for prokaryotic organisms, the process of genome evolution can be studied. Phylogenies for thousands of gene families have now been mapped to the Earth's geological timeline using models of gene gain and loss and horizontal transfer (David & Alm, 2011). This shows that there was a period of innovation of new gene families in the Archaean eon between about 3.3 and 2.8 Ga, which corresponds to a period of diversification among prokaryotic lineages. Several time-calibration points were used. They chose to set the date for LUCA as <3.85 Ga, based on carbon isotope evidence for the earliest life that we discussed above (Mojzsis *et al.* 1996, Rosing 1999). They were also able to follow the change in frequency of genes involved in pathways linked to molecular oxygen, which occurs over the period during which the oxygen level increases in the atmosphere.

Overall, although molecular phylogenetic methods are perhaps not the most reliable way to estimate dates, the molecular studies seem to be consistent and to add some further support to the conclusions derived from geological evidence.

## 7. Discussion - How Quickly Did Life Emerge on Earth?

The timeline that illustrates the essential measurements and the resulting constraints on the emergence of life is detailed in Figure 1. The Earth formed ~24 Myr after the first solids condensed in the Solar System (the latter of which occurred at ~4.568 Ga). The impact of Theia with the proto-Earth ~28–128 Myr after that led to the complete melting of the Earth's surface and the formation of the Moon. Finally, the Earth's magma ocean cooled to form a solid surface, leading to a potentially habitable world another ~0.02–100 Myr later. The absolute earliest the Earth could have been habitable is therefore ~4.5 Ga. On the other hand, the absolute latest the Earth could have been habitable is ~3.9 Ga, as the habitability boundary depends strongly on whether life could emerge during a sustained, declining asteroid bombardment, or survive the potential LHB.

Once life emerged, the first solid imprints of its existence appear in 3.7-Gyr-old rocks of sedimentary origin in a location which is exposed today in southwest Greenland. The imprints of life at 3.7 Ga come in two forms: 1) graphite globules formed out of sedimentary rock which included the remains of organisms with biologically depleted $\delta^{13}$C signatures, and 2) fossilized stromatolite-like structures that formed from photosynthetic organisms in microbial mats. At 3.43–3.48 Ga, more photosynthetic organisms left behind their fossilized imprint in the form of stromatolites. And at 3.4 Ga, the first spheroidal/ellipsoidal, and tubular single-celled organisms imprinted their shapes into the geological record. The rise in atmospheric oxygen due to oxygenic photosynthetic organisms did not begin to leave an imprint in the rock record until 2.95 Ga. This is when oxygenic photosynthesizers likely began to accumulate in shallow marine settings. Finally, at 2.4–2.3 Ga, the oxygen levels in the atmosphere rose in a great oxidation event as oxygenic photosynthesizers proliferated and oxygen sinks near the Earth's surface became saturated.

There is a large range for the outer boundary for the origin of life (4.5–3.9 Ga). Thus in order to analyze this time frame, we will explore two cases. (1) Asteroidal impacts on the Hadean Earth did not postpone or reset the origin of life. (2) Asteroidal impacts continuously postponed the emergence of life, or, specifically in the case of the LHB, reset the entire process. This removes the necessity to debate which scenario for asteroid bombardment occurred on the early Earth, and focuses instead on how asteroid impact frequency may affect the process of forming life.

In case 1, life emerged in the wide time frame of 4.5–3.7 Ga. The outer boundary in this case is based on the earliest estimate for the Moon-forming impact, around 4.516 Ga, followed by efficient radiative cooling to clement temperatures (taking ~0.02–16 Myr). Based on the consolidated evidence in this review, this case represents the widest, and therefore the most conservative time interval for the origin of life. In this case, it would have taken ~800 million years to go from a habitable world to a population of organisms that can imprint their $\delta^{13}$C signature in rocks and form domical stromatolite-like structures. 800 million years is a short period, astronomically speaking, but a long period, evolutionarily speaking. In an RNA world setting (Neveu et al. 2013), life emerged through the Darwinian evolution of chemically produced RNA molecules (Pearce et al. 2017). If this process—beginning with prebiotic chemistry and ending at the LUCA—took 800 million years, perhaps life could be a rare occurrence. After all, 800 million years is 15 % of the time our planet has to be habitable (assuming our oceans boil away due to the greenhouse effect in 1 Gyr (Goldblatt et al. 2013)).

In case 2, life emerged in the shorter time interval of 3.9–3.7 Ga. This is the narrow time frame scenario, which assumes that either the LHB wiped clean any life or pre-life molecules that emerged before 3.9 Ga, or that asteroid impacts were too disruptive for life to emerge during a sustained, declining bombardment period from 4.5–3.9 Ga.

Given these two cases, if life emerged on a timescale of less than 800 million years, does this say anything about ubiquity of life on habitable planets throughout the Universe? In truth we cannot make this conclusion, as the Earth is a sample size of n = 1. In other words, there is a strong selection bias in estimating the probability of life emerging elsewhere in the Universe. Indeed it has been argued that if intelligent life requires a great deal of time to evolve, Earth may be a rare planet, on which life got started unusually early (Carter 1983).

So, is the oldest evidence of life yet to be found? Have all sedimentary rocks older than 3.7 Ga been melted into magma? Have we reached our limit for using $\delta^{13}C$ signatures and need to develop a different technique to find evidence for earlier life? Because biosignatures are detected in the oldest preserved sedimentary rocks, we know that life existed prior to their formation time. And although the 3.8 Ga and 3.85 Ga rock samples from Isua Greenland are not of sedimentary origins, and therefore their low $\delta^{13}C$ content could be explained by high-temperature, abiotic processes (Fedo & Whitehouse 2002), do they still hint at the existence of life at this time? Perhaps an advancement in techniques will ultimately distinguish the low $\delta^{13}C$ value in these samples from the known abiotic mechanisms.

Based on the recent activity in the research fields of biogenic isotopes, stromatolites, and microfossils (Djokic et al. 2017; Dodd et al. 2017; Nutman et al. 2016; Hickman-Lewis et al. 2016; Brasier et al. 2015; Planavsky et al. 2014; Ohtomo et al. 2014; Noffke et al. 2013, Sugitani *et al.* 2013), it would seem that researchers are still actively looking for older evidences of life's existence on Earth. If a high level of interest sustains in these fields, it would not be surprising to see the 3.7 Ga biosignature boundary being pushed to earlier dates in the next decade. However, it is also possible that in the same time, abiotic processes will be revealed to explain both the stromatolite morphologies and the low $\delta^{13}C$ values in old rocks of sedimentary origin.

Whether the LHB occurred has not yet reached consensus. Researchers are still asking whether a single cataclysm occurred (Zellner & Delano 2015), and more evidence will be needed to sway the consensus in either direction. Planet formation theorists are actively improving models for the formation of our solar system (Jacobson et al. 2014; Raymond & Morbidelli 2014), which could eventually constrain the inwards and outwards migration of Jupiter to better understand the likelihood of a late lunar cataclysm being caused by such an event.

Finally, many researchers such as Rugheimer et al. (2015), Mollière et al. (2015), Hu & Seager (2014), Kaltenegger et al. (2013) are improving atmospheric models, which may someday be sophisticated enough to include $CO_2$ subduction and degassing from meteorites. These models could better constrain the cooling time for the Earth's magma mantle, and the likelihood of the Earth losing all of its atmosphere and freezing over.

As a final thought, we note that even the shorter 200-million-year time frame is very much longer than the life times of individual biomolecules. A key step for the origin of life is the creation of an autocatalytic, self-replicating chemical system. We envisage a 'searching and waiting period' in which chemical synthesis generates and explores a diversity of random, non-functional biopolymer sequences, until a rare event occurs that establishes a stable small group of replicators. Our understanding of molecular evolution, the nature of the earliest replicators and the environment in which they existed is not good enough to make a first-principles prediction of how long a search period would have been required. The time interval between the habitability and biosignature boundaries, which we have been considering in this paper, tells us some information about how long the search period might have been, but it should be remembered that the time resolution of our biological and geochemical records is poor, and there remains the possibility that the search and wait period could have been much shorter than 200 million years. The initiation of life may be considered as a rare event on a biochemical and evolutionary timescale, and at the same time very rapid on the scale of planetary lifetimes. A key goal of the field of astrobiology and the origin of life is to estimate the average searching and waiting period on habitable planets, and hence to be able to give reliable estimates of the frequency of life on other worlds.

## Acknowledgments

We are grateful to two anonymous referees whose reports helped considerably to clarify the text. We are especially grateful to Allyson Brady, Greg Slater, and Radhey Gupta for their invaluable comments on the various topics in this review. We thank Steve Janzen - at McMaster Media Resources - for his superb work in rendering Figure 1. This work was supported by the NSERC discovery grants of P.G.H. and R.E.P. Finally, B.K.D.P was supported by a NSERC Canada graduate scholarship (CGS-M) and an Ontario graduate scholarship (OGS).

## Author Disclosure Statement

No competing financial interests exist. Final publication is available from Mary Ann Liebert, Inc., publishers http://dx.doi.org/10.1089/ast.2017.1674